\documentclass{pasj00}
\draft

\begin{document}
\SetRunningHead{Seta et al.}{Suzaku and multi-wavelength observation of OJ~287}
\Received{2009/00/00}
\Accepted{2009/00/00}

\title{Suzaku and Multi-wavelength Observations of OJ~287 during the Periodic Optical Outburst in 2007} 

\author{
Hiromi    \textsc{Seta},                \altaffilmark{1}
Naoki     \textsc{Isobe},               \altaffilmark{2}
Makoto~S. \textsc{Tashiro},             \altaffilmark{1}
Yuichi    \textsc{Yaji},                \altaffilmark{1}\\
Akira     \textsc{Arai},                \altaffilmark{3}
Masayuki  \textsc{Fukuhara},            \altaffilmark{4,5}
Kotaro    \textsc{Kohno},               \altaffilmark{6} 
Koichiro  \textsc{Nakanishi},           \altaffilmark{5}\\
Mahito    \textsc{Sasada},              \altaffilmark{3}
Yoshito   \textsc{Shimajiri},           \altaffilmark{4,5}
Tomoka    \textsc{Tosaki},              \altaffilmark{5}
Makoto    \textsc{Uemura},              \altaffilmark{7}
\and 
H         \textsc{Anderhub},            \altaffilmark{8}
L~A.      \textsc{Antonelli},           \altaffilmark{9}
P         \textsc{Antoranz},            \altaffilmark{10}
M         \textsc{Backes},              \altaffilmark{11}
C         \textsc{Baixeras},            \altaffilmark{12}\\
S         \textsc{Balestra},            \altaffilmark{10}
J~A.      \textsc{Barrio},              \altaffilmark{10}
D         \textsc{Bastieri},            \altaffilmark{13}
J         \textsc{Becerra Gonz\'alez},  \altaffilmark{14}
J~K.      \textsc{Becker},              \altaffilmark{11}\\
W         \textsc{Bednarek},            \altaffilmark{15}
K         \textsc{Berger},              \altaffilmark{15}
E         \textsc{Bernardini},          \altaffilmark{16}
A         \textsc{Biland},              \altaffilmark{8}
R~K.      \textsc{Bock},                \altaffilmark{17,13}\\
G         \textsc{Bonnoli},             \altaffilmark{18}
P         \textsc{Bordas},              \altaffilmark{19}
D         \textsc{Borla Tridon},        \altaffilmark{17}
V         \textsc{Bosch-Ramon},         \altaffilmark{19}
D         \textsc{Bose},                \altaffilmark{10}\\
I         \textsc{Braun},               \altaffilmark{8}
T         \textsc{Bretz},               \altaffilmark{20}
I         \textsc{Britvitch},           \altaffilmark{8}
M         \textsc{Camara},              \altaffilmark{10}
E         \textsc{Carmona},             \altaffilmark{17}
S         \textsc{Commichau},           \altaffilmark{8}\\
J~L.      \textsc{Contreras},           \altaffilmark{10}
J         \textsc{Cortina},             \altaffilmark{21}
M~T.      \textsc{Costado},             \altaffilmark{14,22}
S         \textsc{Covino},              \altaffilmark{9}
V         \textsc{Curtef},              \altaffilmark{11}\\
F         \textsc{Dazzi},               \altaffilmark{23,35}
A.        \textsc{De Angelis},          \altaffilmark{23}
E         \textsc{De Cea del Pozo},     \altaffilmark{24}
R         \textsc{de los Reyes},        \altaffilmark{10}\\
B         \textsc{De Lotto},            \altaffilmark{23}
M         \textsc{De Maria},            \altaffilmark{23}
F         \textsc{De Sabata},           \altaffilmark{23}
C         \textsc{Delgado Mendez},      \altaffilmark{14,32}\\
A         \textsc{Dominguez},           \altaffilmark{25}
D         \textsc{Dorner},              \altaffilmark{8}
M         \textsc{Doro},                \altaffilmark{13}
D         \textsc{Elsaesser},           \altaffilmark{20}
M         \textsc{Errando},             \altaffilmark{21}\\
D         \textsc{Ferenc},              \altaffilmark{26}
E         \textsc{Fern\'andez},         \altaffilmark{21}
R         \textsc{Firpo},               \altaffilmark{21}
M~V.      \textsc{Fonseca},             \altaffilmark{10}
L         \textsc{Font},                \altaffilmark{12}
N         \textsc{Galante},             \altaffilmark{17}\\
R~J.      \textsc{Garc\'{\i}a L\'opez}, \altaffilmark{14,22}
M         \textsc{Garczarczyk},         \altaffilmark{21}
M         \textsc{Gaug},                \altaffilmark{14}
F         \textsc{Goebel},              \altaffilmark{17,36}
D         \textsc{Hadasch},             \altaffilmark{12}\\
M         \textsc{Hayashida},           \altaffilmark{17,33}
A         \textsc{Herrero},             \altaffilmark{14,22}
D         \textsc{Hildebrand},          \altaffilmark{8}
D         \textsc{H\"ohne-M\"onch},     \altaffilmark{20}
J         \textsc{Hose},                \altaffilmark{17}\\
C~C.      \textsc{Hsu},                 \altaffilmark{17}
T         \textsc{Jogler},              \altaffilmark{17}
D         \textsc{Kranich},             \altaffilmark{8}
A         \textsc{La Barbera},          \altaffilmark{9}
A         \textsc{Laille},              \altaffilmark{26}
E         \textsc{Leonardo},            \altaffilmark{18}\\
E         \textsc{Lindfors},            \altaffilmark{27}
S         \textsc{Lombardi},            \altaffilmark{13}
F         \textsc{Longo},               \altaffilmark{23}
M         \textsc{L\'opez},             \altaffilmark{13}
E         \textsc{Lorenz},              \altaffilmark{8,17}
P         \textsc{Majumdar},            \altaffilmark{16}\\
G         \textsc{Maneva},              \altaffilmark{28}
N         \textsc{Mankuzhiyil},         \altaffilmark{23}
K         \textsc{Mannheim},            \altaffilmark{20}
L         \textsc{Maraschi},            \altaffilmark{9}
M         \textsc{Mariotti},            \altaffilmark{13}\\
M         \textsc{Mart\'{\i}nez},       \altaffilmark{21}
D         \textsc{Mazin},               \altaffilmark{21}
M         \textsc{Meucci},              \altaffilmark{18}
M         \textsc{Meyer},               \altaffilmark{20}
J~M.      \textsc{Miranda},             \altaffilmark{10}\\
R         \textsc{Mirzoyan},            \altaffilmark{17}
H         \textsc{Miyamoto},            \altaffilmark{17}
J         \textsc{Mold\'on},            \altaffilmark{19}
M         \textsc{Moles},               \altaffilmark{25}
A         \textsc{Moralejo},            \altaffilmark{21}
D         \textsc{Nieto},               \altaffilmark{10}\\
K         \textsc{Nilsson},             \altaffilmark{27}
J         \textsc{Ninkovic},            \altaffilmark{17}
N         \textsc{Otte},                \altaffilmark{17,34}
I         \textsc{Oya},                 \altaffilmark{10}
R         \textsc{Paoletti},            \altaffilmark{18}
J~M.      \textsc{Paredes},             \altaffilmark{19}\\
M         \textsc{Pasanen},             \altaffilmark{27}
D         \textsc{Pascoli},             \altaffilmark{13}
F         \textsc{Pauss},               \altaffilmark{8}
R~G.      \textsc{Pegna},               \altaffilmark{18}
M~A.      \textsc{Perez-Torres},        \altaffilmark{25}\\
M         \textsc{Persic},              \altaffilmark{23,29}
L         \textsc{Peruzzo},             \altaffilmark{13}
F         \textsc{Prada},               \altaffilmark{25}
E         \textsc{Prandini},            \altaffilmark{13}
N         \textsc{Puchades},            \altaffilmark{21}\\
I         \textsc{Reichardt},           \altaffilmark{21}
W         \textsc{Rhode},               \altaffilmark{11}
M         \textsc{Rib\'o},              \altaffilmark{19}
J         \textsc{Rico},                \altaffilmark{30,21}
M         \textsc{Rissi},               \altaffilmark{8}
A         \textsc{Robert},              \altaffilmark{12}\\
S         \textsc{R\"ugamer},           \altaffilmark{20}
A         \textsc{Saggion},             \altaffilmark{13}
T~Y.      \textsc{Saito},               \altaffilmark{17}
M         \textsc{Salvati},             \altaffilmark{9}
M         \textsc{Sanchez-Conde},       \altaffilmark{25}\\
K         \textsc{Satalecka},           \altaffilmark{16}
V         \textsc{Scalzotto},           \altaffilmark{13}
V         \textsc{Scapin},              \altaffilmark{23}
T         \textsc{Schweizer},           \altaffilmark{17}
M         \textsc{Shayduk},             \altaffilmark{17}\\
S~N.      \textsc{Shore},               \altaffilmark{31}
N         \textsc{Sidro},               \altaffilmark{21}
A         \textsc{Sierpowska-Bartosik}, \altaffilmark{24}
A         \textsc{Sillanp\"a\"a},       \altaffilmark{27}
J         \textsc{Sitarek},             \altaffilmark{17,15}\\
D         \textsc{Sobczynska},          \altaffilmark{15}
F         \textsc{Spanier},             \altaffilmark{20}
A         \textsc{Stamerra},            \altaffilmark{18}
L~S.      \textsc{Stark},               \altaffilmark{8}
L         \textsc{Takalo},              \altaffilmark{27}\\
F         \textsc{Tavecchio},           \altaffilmark{9}
P         \textsc{Temnikov},            \altaffilmark{28}
D         \textsc{Tescaro},             \altaffilmark{21}
M         \textsc{Teshima},             \altaffilmark{17}
M         \textsc{Tluczykont},          \altaffilmark{16}\\
D~F.      \textsc{Torres},              \altaffilmark{30,24}
N         \textsc{Turini},              \altaffilmark{18}
H         \textsc{Vankov},              \altaffilmark{28}
R~M.      \textsc{Wagner},              \altaffilmark{17}
W         \textsc{Wittek},              \altaffilmark{17}\\
V         \textsc{Zabalza},             \altaffilmark{19}
F         \textsc{Zandanel},            \altaffilmark{25}
R         \textsc{Zanin},               \altaffilmark{21}
J         \textsc{Zapatero}             \altaffilmark{12}\\
(The MAGIC Collaboration)
}

\altaffiltext{1} {Department of Physics, Saitama University, 255 Shimo-Okubo, Sakura, Saitama 338-8570}
\email{seta@heal.phy.saitama-u.ac.jp}
\altaffiltext{2} {Department of Astronomy, Kyoto University, Kitashirakawa-Oiwakecho, Sakyo-ku, Kyoto 606-8502}
\altaffiltext{3} {Department of Physical Science, Hiroshima University, 1-3-1 Kagamiyama, Higashi-Hiroshima 739-8526}
\altaffiltext{4} {Department of Astronomy School of Science, University of Tokyo, Bunkyo, Tokyo 113-0033}
\altaffiltext{5} {Nobeyama Radio Observatory, Minamimaki, Minamisaku, Nagano 384-1305}
\altaffiltext{6} {Institute of Astronomy, The University of Tokyo, 2-21-1 Osawa, Mitaka, Tokyo 181-0015}
\altaffiltext{7} {Astrophysical Science Center, Hiroshima University, 1-3-1 Kagamiyama, Higashi-Hiroshima 739-8526}
\altaffiltext{8} {ETH Zurich, CH-8093 Switzerland}
\altaffiltext{9} {INAF National Institute for Astrophysics, I-00136 Rome, Italy}
\altaffiltext{10} {Universidad Complutense, E-28040 Madrid, Spain}
\altaffiltext{11} {Technische Universit\"at Dortmund, D-44221 Dortmund, Germany}
\altaffiltext{12} {Universitat Aut\`onoma de Barcelona, E-08193 Bellaterra, Spain}
\altaffiltext{13} {Universit\`a di Padova and INFN, I-35131 Padova, Italy}
\altaffiltext{14} {Inst. de Astrof\'{\i}sica de Canarias, E-38200 La Laguna, Tenerife, Spain}
\altaffiltext{15} {University of \L\'od\'z, PL-90236 Lodz, Poland}
\altaffiltext{16} {Deutsches Elektronen-Synchrotron (DESY), D-15738 Zeuthen, Germany}
\altaffiltext{17} {Max-Planck-Institut f\"ur Physik, D-80805 M\"unchen, Germany}
\altaffiltext{18} {Universit\`a  di Siena, and INFN Pisa, I-53100 Siena, Italy}
\altaffiltext{19} {Universitat de Barcelona (ICC/IEEC), E-08028 Barcelona, Spain}
\altaffiltext{20} {Universit\"at W\"urzburg, D-97074 W\"urzburg, Germany}
\altaffiltext{21} {IFAE, Edifici Cn., Campus UAB, E-08193 Bellaterra, Spain}
\altaffiltext{22} {Depto. de Astrofisica, Universidad, E-38206 La Laguna, Tenerife, Spain}
\altaffiltext{23} {Universit\`a di Udine, and INFN Trieste, I-33100 Udine, Italy}
\altaffiltext{24} {Institut de Cienci\`es de l'Espai (IEEC-CSIC), E-08193 Bellaterra, Spain}
\altaffiltext{25} {Inst. de Astrof\'{\i}sica de Andalucia (CSIC), E-18080 Granada, Spain}
\altaffiltext{26} {University of California, Davis, CA-95616-8677, USA}
\altaffiltext{27} {Tuorla Observatory, University of Turku, FI-21500 Piikki\"o, Finland}
\altaffiltext{28} {Inst. for Nucl. Research and Nucl. Energy, BG-1784 Sofia, Bulgaria}
\altaffiltext{29} {INAF/Osservatorio Astronomico and INFN, I-34143 Trieste, Italy}
\altaffiltext{30} {ICREA, E-08010 Barcelona, Spain}
\altaffiltext{31} {Universit\`a  di Pisa, and INFN Pisa, I-56126 Pisa, Italy}
\altaffiltext{32} {now at: Centro de Investigaciones Energ ticas, Medioambientales y Tecnol gicas (CIEMAT), Madrid, Spain}
\altaffiltext{33} {now at: SLAC National Accelerator Laboratory and KIPAC, CA, 94025, USA}
\email{mahaya@slac.stanford.edu}
\altaffiltext{34} {now at: University of California, Santa Cruz, CA 95064, USA}
\altaffiltext{35} {supported by INFN Padova}
\altaffiltext{36} {deceased}

\maketitle

\begin{abstract}
Suzaku observations of the blazar OJ~287 
were performed in 2007 April 10 -- 13 and November 7 -- 9. 
They correspond to a quiescent and a flaring state, respectively.
The X-ray spectra of the source can be 
well described with 
single power-law models in both exposures.  
The derived X-ray photon index and the flux density at $1$ keV 
were found to be 
$\Gamma = 1.65 \pm 0.02$ and $S_{\rm 1 keV} = 215 \pm 5$~nJy,
in the quiescent state.
In the flaring 
state, the source exhibited a harder X-ray spectrum
($\Gamma = 1.50 \pm 0.01$) 
with a nearly doubled X-ray flux density $S_{\rm 1 keV} = 404^{+6}_{-5}$~nJy.
Moreover, significant hard X-ray signals were detected up to $\sim 27$ keV.
In cooperation with the Suzaku, 
simultaneous radio, optical, and very-high-energy $\gamma$-ray observations 
of OJ~287 were performed with the Nobeyama Millimeter Array, 
the KANATA telescope, and the MAGIC telescope, respectively.
The radio and optical fluxes in the flaring 
state
($3.04 \pm 0.46$~Jy and $8.93 \pm 0.05$~mJy at 86.75~Hz and in the $V$-band, 
respectively) 
were found to be higher
by a factor of $2$ -- $3$ 
than those in the quiescent state 
($1.73 \pm 0.26$~Jy and $3.03 \pm 0.01$~mJy at 86.75~Hz and in the $V$-band, 
respectively) . 
No notable $\gamma$-ray events were detected 
in either observation. 
The spectral energy distribution of OJ~287 indicated that 
the X-ray spectrum was dominated by inverse Compton radiation
in both observations,
while synchrotron radiation exhibited a spectral cutoff around the optical frequency. 
Furthermore, no significant difference in 
the synchrotron cutoff frequency was found 
between the quiescent and flaring states.
According to a simple synchrotron self-Compton model, 
the change of the spectral energy distribution is due to 
an increase in the energy density of electrons
with small changes 
of both the magnetic field strength and
the maximum Lorentz factor of electrons.
\end{abstract}

\section{Introduction}
At a redshift of $z = 0.306$ \citep{Stickel}, 
OJ~287 is one of the archetypal and most studied blazars. 
An outstanding characteristic of the object is 
its recurrent optical outbursts with a period of $11.65$ years, 
as revealed by optical data 
spanning more than $100$ years \citep{SMBH}.
The outburst in $1994$ motivated 
a worldwide multi-wavelength observation campaign named 
``The OJ~$94$ project'' \citep{OJ94}.
This project 
confirmed the periodicity and revealed
that the optical outbursts consist of two peaks corresponding to flares 
with an interval of about one year \citep{double_peak}. 
OJ~287 is suggested to be a binary black hole system in which a secondary
black hole pierces the accretion disk of the primary black hole and
produces two impact flushes per period \citep{Valtonen_nature}.
Interestingly, the first flare exhibited 
a low radio flux with decreasing radio polarization 
and a relatively short duration (a few months),
while the second one had a high radio flux with increasing radio polarization,
and lasted about half a year \citep{double_peak,disk-jet,Pursimo}. 
The differences between the two flares may be interpreted as 
the first flare have a thermal origin in the vicinity of the black hole and the accretion disk,
while the second one originate from synchrotron radiation from the jet \citep{disk-jet},
although we have not yet obtained any convincing evidence supporting this interpretation. 

The multi-wavelength spectral energy distribution (SED) has 
the potential to resolve the physical state of OJ~287 during the flares.
In general, the SED of blazars is characterized 
by two broad humps~(e.g., \cite{Fos98}); 
the low-energy component, 
with wavelengths in the range between radio to ultraviolet and X-ray, 
is widely regarded as synchrotron radiation (SR) from relativistic electrons within the jet.
The high-energy component, with wavelengths in the range between X-rays and $\gamma$-rays, 
is interpreted as inverse-Compton (IC) scattering. 
In one of the simple emission models, named 
``synchrotron self-Compton (SSC) model'', 
relativistic electrons scatter SR photons produced by the same population of electrons (e.g., \cite{Ghi98}).
For low-frequency peaked BL Lac objects (LBLs), a class of blazars to
which OJ~287 belongs, 
the SR peak is located in the range between sub-mm and optical wavelengths~\citep{Pad95}.
IC scattering in LBLs can emit radiation up to very-high-energy (VHE: E$>$50 GeV) $\gamma$-rays during their optical high states; 
VHE $\gamma$-ray emission has been detected, for example, from BL Lacertae~\citep{BLLac} and S5~0716+714~\citep{Tes08}.
These two components usually intersect with each other in the X-ray band (e.g., \cite{S50716}).
Therefore, the X-ray spectrum of LBLs  
should be highly sensitive to the change
of the magnetic field ($B$) and/or the maximum Lorentz factor 
($\gamma_{\rm max}$) of the electrons,
since the SR peak scales as $\propto B \gamma_{\rm max}^2$. 
In fact, the X-ray spectrum of the object, 
obtained in previous observations, 
can be successfully interpreted on the basis of the interplay 
between the SR and the IC component \citep{Idesawa,Isobe}.
However, all of these X-ray observations were carried out during the first flare,
and there is no information regarding the X-ray spectrum 
of OJ~287 in the second flare. 

In the period between $2005$ and $2008$,
OJ~287 was predicted to move to the last 
active phase, 
and was in fact reported to exhibit the first optical outburst 
in $2005$ November \citep{Valtonen_nature,OJ287_2005_2}.
Since the second flare of the source was expected to be in the fall of $2007$ 
(\cite{Valtonen2,Kidger}),
we organized two X-ray and simultaneous multi-wavelength observations, 
in the quiescent state (MWL~I) between the two outbursts 
and in the second flaring state (MWL~II), with the
objective to reveal the characteristics of the second flare,
in comparison with the quiescent state.
Based on the results, we provide a discussion regarding the differences 
between the first and the second flare.
The Suzaku X-ray observation in MWL~I was conducted 
during $2007$ April $10$ -- $13$,
when the source was optically quiescent with an $R$-band magnitude of
about $15$.
We triggered the MWL~II Suzaku observation in $2007$ November $7$,
on condition that the object remained 
brighter than $14$-th magnitude for more than one week,
as indicated by the optical monitoring data of the source
taken at the Tuorla Observatory and the KVA observatory 
\footnote{http://users.utu.fi/kani/1m/OJ\_287\_jy.html}.
We also conducted monitoring observations at
radio, optical, and VHE $\gamma$-ray frequencies
with the Nobeyama Millimeter Array (NMA), the KANATA telescope, 
and the MAGIC telescope, respectively.

The X-ray observations and the spectral results are presented in section 2. 
section 3 shows the radio, optical, and VHE $\gamma$-ray results.
These are followed by the discussion on the physical state of the second flare
in section 4 on the basis of SED obtained in these multi-frequency observations.

\section{X-ray Observations and Results}
\subsection{Observations and data reduction}
\label{sec:reduction}
The Suzaku pointing observation for MWL~I was conducted 
between 19:47:00 UT 2007 April 10 
and 11:10:19 UT April 13 (ObsID 702009010),
and the pointing for MWL~II was conducted 
between 11:24:00 UT 2007 November 7 
and 21:30:23 November 9 (ObsID 702008010).
The X-ray Imaging Spectrometer (XIS; \cite{Koyama})
and the Hard X-ray Detector (HXD; \cite{Takahashi}; \cite{Kokubun})
onboard Suzaku were operated in the normal clocking mode 
with no window option, and in the normal mode, respectively.  

We placed OJ~287 at the HXD-nominal position \citep{Serlemitosos}.
We performed data reduction by 
using the HEADAS~$6.5.1$ software package
and referring to the calibration data base (CALDB) of 
the XIS, the X-ray telescope (XRT; \cite{Serlemitosos}),
and the HXD 
as of 2008 September 5, 2008 July 9, and 2008 August 11, respectively. 

We reprocessed the XIS data in MWL~I
since the version of the standard processing (Revision~$2.0.6.13$) was obsolete, 
whereas we utilized the pipeline cleaned events 
from the Revision~$2.1.6.16$ processing for MWL~II. 
The reprocessed/cleaned XIS data were filtered 
under the following criteria;
the spacecraft is outside the south Atlantic anomaly (SAA), 
the time interval after an exit from the SAA is longer than $436$ s,
the geomagnetic cutoff rigidity (COR) is higher than $6$ GV, 
the source elevation above the rim of bright and night Earth (ELV) is
higher than \timeform{20D} and \timeform{5D}, respectively, 
and the XIS data are free from telemetry saturation. 
These procedures yielded $85.3$ ks and $102.6$ ks of good exposures,
for MWL~I and MWL~II. 
In the scientific analysis below,
we utilize only the events with grades of $0$, $2$, $3$, $4$, or $6$.

We reprocessed the HXD data in both observations by using the latest gain
file in the CALDB. 
The HXD data were screened under the following criteria; 
the time interval before and after the SAA is longer than $180$ s and $500$ s,
COR and ELV are higher than $6$~GV and \timeform{5D}, respectively.
As a result, we obtained $93.6$ ks and $102.9$ ks of good exposures,      
for MWL~I and MWL~II. 

\subsection{XIS Results}
\label{sec:XIS}
\begin{figure}[ht]
\begin{center}
\FigureFile(150mm,150mm){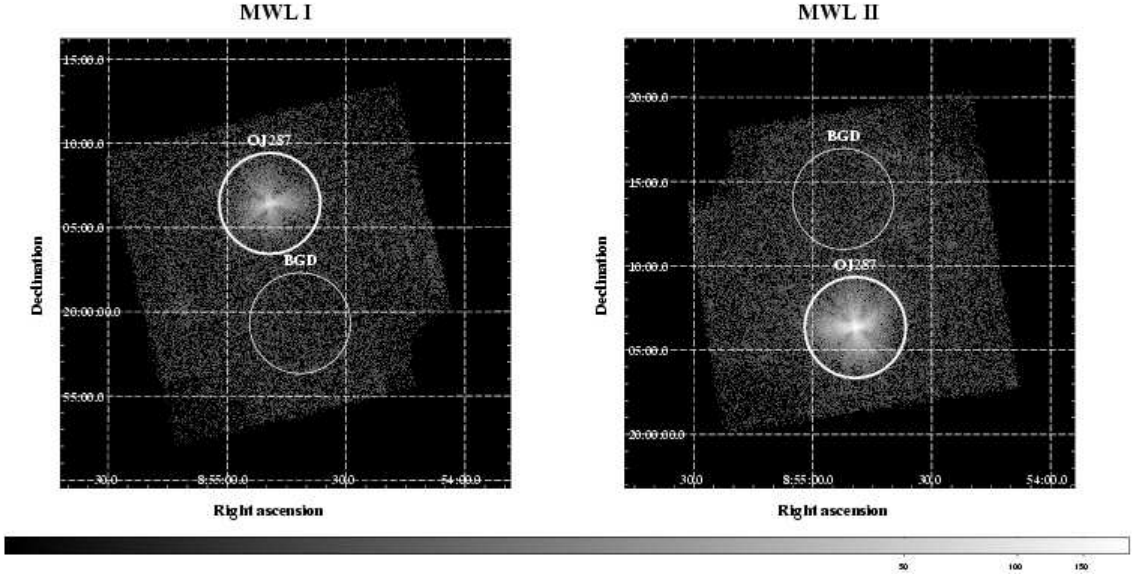}
\end{center}
\caption{
The $0.5$ -- $10$ keV XIS image of OJ~287 in MWL~I (left) and MWL~II (right).
Data from all XIS CCD chips (XIS~$0$, $1$, and $3$) were summed up. 
The region of the calibration sources was removed. 
Background subtraction and exposure correction 
are not applied to the image.
Both panels are drawn in the same grayscale.
The source and the background signals are integrated 
within the solid and dashed circles, respectively.
}
\label{fig:xis_image}
\end{figure}

Figure~\ref{fig:xis_image} shows the $0.5$ -- $10$ keV XIS images in
MWL~I and MWL~II.
OJ~287 was clearly detected at the position of 
($\alpha, \delta$) = (\timeform{08h54m48s.87}, \timeform{20D06'30''.6})
in J2000.0 coordinates,
with no other apparent contaminating source within the XIS field of view.
The image indicates  
that the intensity of the source was higher in MWL~II than in MWL~I.
We accumulated the source signals within
the solid circles with a radius of \timeform{3'} in figure~\ref{fig:xis_image}. 
The background events were integrated within the same radius (dashed circles) 
at symmetric positions around the optical axis of the XRT. 

\begin{figure}[ht]
\begin{center}
\FigureFile(80mm,80mm){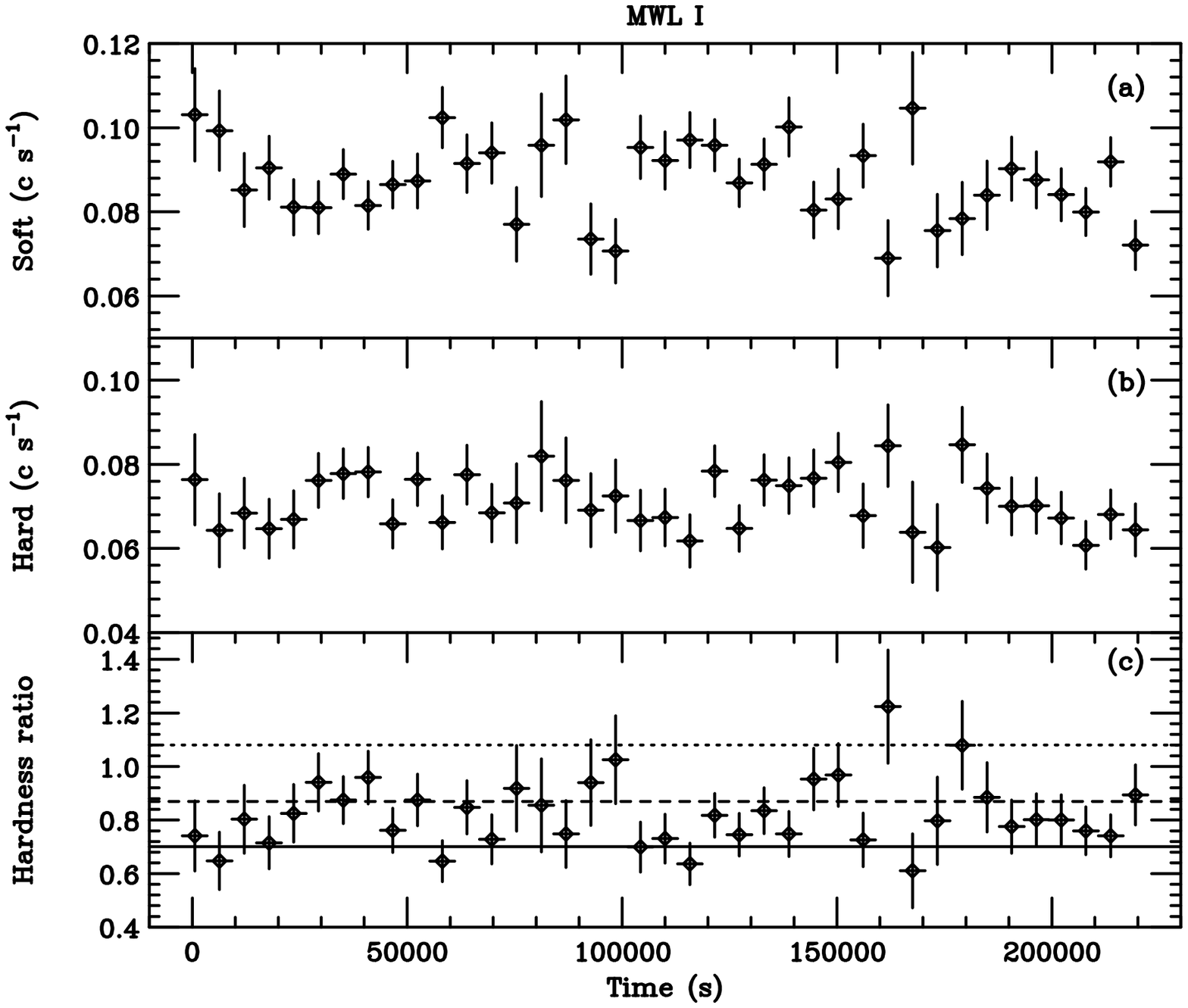}
\FigureFile(80mm,80mm){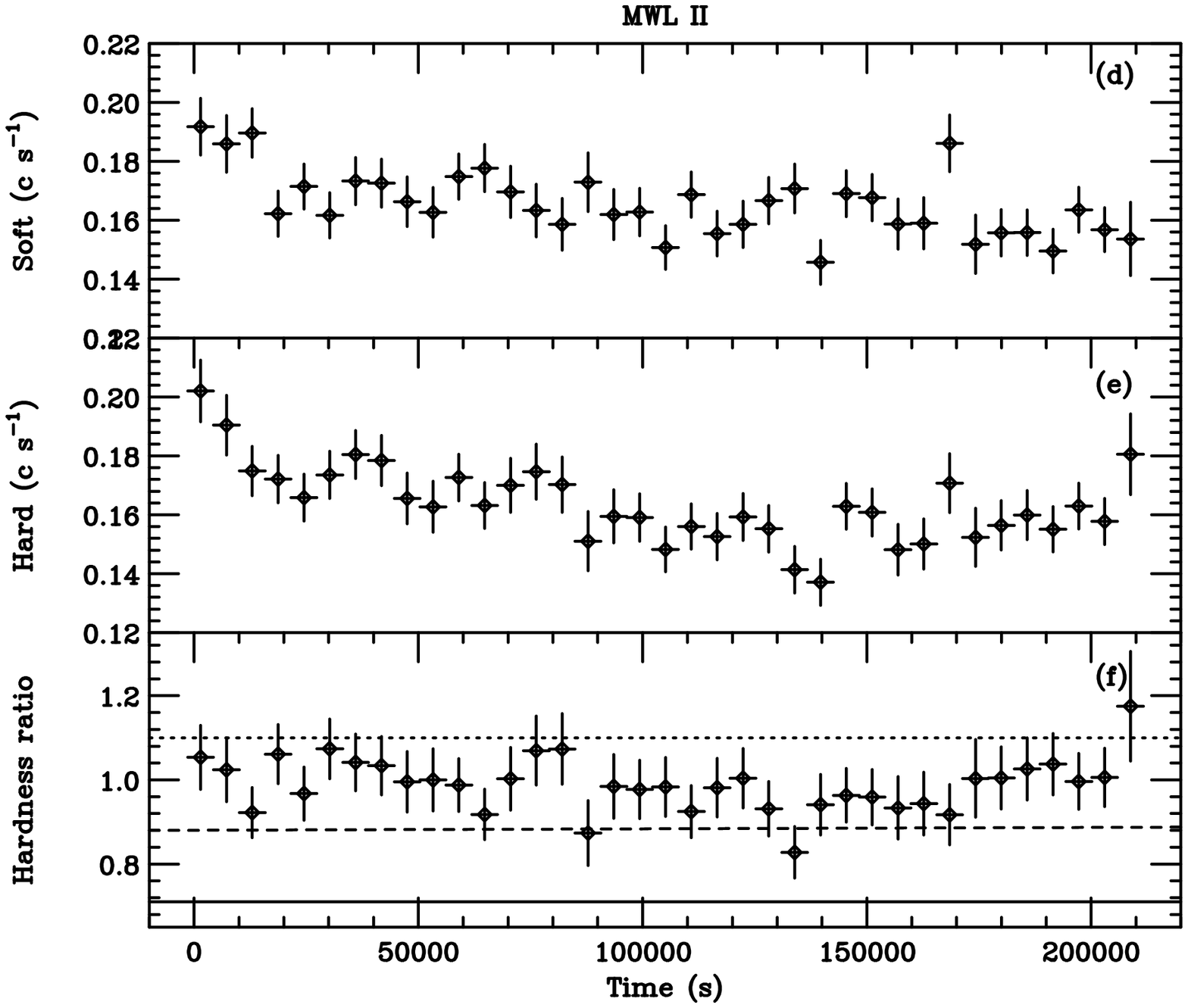}
\end{center}
\caption{
The XIS FI lightcurves of OJ~287 obtained in MWL~I and MWL~II. 
The time bin was set to $5760$~s.
Panels (a) and (d) show the soft energy band lightcurves ($0.5$ -- $2$ keV), while
panels (b) and (e) show the medium energy band lightcurves ($2$ -- $10$ keV)
in each observation.
The hardness, which is simply calculated as the ratio of the hard band count rate
to the soft band one, is shown in panels (c) and (f).
The dotted, the dashed, and the solid lines in panels (c) and (f) indicate
the predictions provided by the PL model with 
$\Gamma$=$1.4$, $1.6$, and $1.8$, respectively.
}
\label{fig:XIS_lc}
\end{figure}

In figure \ref{fig:XIS_lc}, 
we show the background-subtracted X-ray lightcurves of OJ~287 in MWL~I and MWL~II. 
The data from the two front-illuminated (FI) CCD cameras (XIS~$0$ and $3$; \cite{Koyama}) were co-added.
In MWL~I, 
the time-averaged count rates in the soft ($0.5$ -- $2$ keV) 
and medium ($2$ -- $10$ keV) bands were measured to be
$0.088 \pm 0.001$ cts s$^{-1}$ and 
$0.071 \pm 0.001$ cts s$^{-1}$, respectively, 
while they nearly doubled to  
$0.166 \pm 0.001$ cts s$^{-1}$ and 
$0.164 \pm 0.002$ cts s$^{-1}$ in MWL~II. 
With $\chi^2 = 42.9$ and $27.7$ for 39 degrees of freedom (d.o.f.),
the lightcurve indicates no significant variation 
during MWL~I 
in either energy band.
On the other hand, in MWL~II, we found that 
the source flux gradually decreased by a factor of $1.3$ 
in the first half of the observation ($\sim 1.5$ days; 
$\chi^2/{\rm d.o.f.} = 58.5/37$ and $71.3/37$ in the soft and medium bands).
However, the hardness ratio, 
which was simply calculated by dividing the count rate of the medium band by that of the soft band,
indicates no significant spectral variation 
in either observation. 
Therefore, we evaluated the averaged spectra as presented below.

Figure~\ref{fig:XIS_spec} shows the background-subtracted XIS spectra of
OJ~287, 
without removing the instrumental response. In this case,
significant X-ray signals were detected 
in the range of $0.5$ -- $10$ keV and $0.4$ -- $8$ keV,
with the FI CCDs and the backside-illuminated (BI) CCD (XIS~$1$).
The spectra appear to be featureless, 
without any absorption or emission lines.

We fitted the spectra with a single power-law (PL) model
modified for photoelectric absorption and subsequently
calculated the response matrix function and the
auxiliary response file by
using {\tt xisrmfgen} and {\tt xissimarfgen} 
\citep {Ishisaki}, respectively.
The effects of absorption caused by the presence of contaminants on the surface of the optical blocking
filter of the CCD 
is taken into account in {\tt xissimarfgen}. 
The FI and BI spectra were jointly fitted
by allowing deviations between their respective model normalizations.
Because we found that the fluxes from the BI and FI spectra were
in good agreement within 7~\%, we adopt the FI value in the present paper.
We fixed the absorption column density at the Galactic value
($N_{\rm H} = 2.56 \times 10^{20}$ cm$^{-2}$; \cite{Kalberla}).
The PL model became acceptable, yielding the best-fit photon index of 
$\Gamma = 1.65 \pm 0.02$ and $\Gamma = 1.50 \pm 0.01$ 
for MWL~I and MWL~II, respectively. 
Thus, we found that OJ~287 showed a harder X-ray spectrum 
in MWL~II. 
As indicated in the lightcurve (figure~\ref{fig:XIS_lc}),
the flux density of the source in MWL~II ($404^{+6}_{-5}$ nJy) 
was higher than that in MWL~I ($215 \pm 5$ nJy) by a factor of $2$. 

\subsection{HXD-PIN Results}
It is crucial to evaluate the non-X-ray background (NXB), 
before examining the hard X-ray spectrum observed with the HXD-PIN
(\cite{Takahashi}; \cite{Kokubun}). 
Although the NXB varies under various cosmic ray environments in the
orbit, the HXD team revealed a set of control parameters for reproducing the NXB
and supplied simulated NXB event files for each observation.
The latest version of the NXB model, named ``tuned''-NXB (LCFITDT in \cite{Fukazawa}), 
was reported to have
a systematic reproducibility error of $2.31$~$\%$ and of $0.99$~$\%$ 
with $10$~ks and $40$~ks exposures, respectively, in the $15$ -- $40$~keV band.

\begin{figure}[ht]
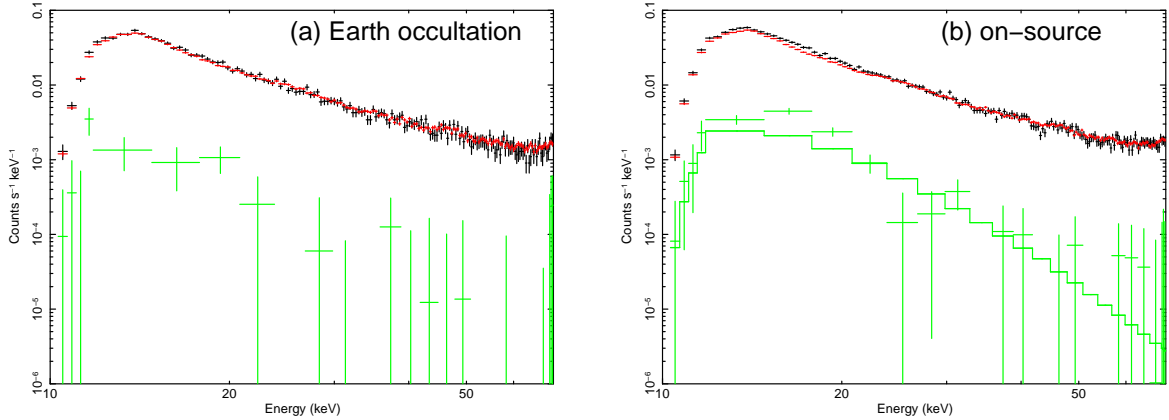

\begin{center}
\FigureFile(80mm,80mm){figure4a.ps}
\FigureFile(80mm,80mm){figure4b.ps}
\end{center}
\caption{
The HXD-PIN spectra of OJ~287 in MWL~I,
obtained in the Earth occultation (panel~a)
and the on-source observations (panel~b).
The black, red, and green data points indicate
the data, NXB model, and NXB-subtracted spectra, respectively.
The green histogram in panel (a) shows
the CXB model spectrum (see the text).
}
\label{fig:PIN_qui}
\end{figure}

First, we analyzed the HXD-PIN data in MWL~I. 
Since the data during Earth occultation is 
essentially dominated by the NXB events, 
these data allow for 
the reproducibility of the NXB model to be evaluated.
The data were obtained with the same criteria as in the case of the on-source observation (section~\ref{sec:reduction}),
with the exception that ELV $<$ \timeform{-5D} instead of ELV $>$ \timeform{5D}.
The exposure of 44.1 ks was attained under these conditions. 
Figure \ref{fig:PIN_qui} (a) compares the observed Earth-occultated (black), 
the NXB model (red), and the NXB-subtracted (green) spectra. 
Table~\ref{table:HXD_log_MWLI} summarizes the statistics of the data and
the NXB model in MWL~I.
We found a data excess of $3.8 \pm 1.0~\%$ over the NXB model below 20~keV.
The apparent excess below 20~keV is within a $3~\sigma$ level of the
current uncertainty of the NXB model (\cite{Fukazawa}).
Therefore, we consider that the excess is 
an artifact produced by fluctuations in the NXB model.
In other words, 
the NXB model underestimates the real NXB data with $\sim 4$~\% in this
observation.
It is necessary to take this into account for the on-source data.

In a similar manner,
we compared the on-source and NXB model spectra, and the results are shown in figure \ref{fig:PIN_qui} (b).
We found a significant data excess over the NXB model, in the 12 -- 40 keV range. 
This excess includes not only the source signals,
but also those of the cosmic X-ray background (CXB). 
In order to evaluate the CXB component, 
we introduced the CXB spectrum within the HXD field of view as  
\begin{displaymath}
 F(E) = (9.412 \times 10^{-3}) \biggl(\frac{E}{1\rm{keV}} \biggl)^{-1.29}
\exp  \biggl[\frac{-E}{40 \rm{keV}}  \biggl]~\rm{photons}~s^{-1} cm^{-2}
 \rm{keV}^{-1}, 
\end{displaymath}
in accordance to \citet{bolt}. 
We folded this spectrum with 
the HXD-PIN response of {\tt ae\_hxd\_pinhxnome3\_20080129.rsp} which was 
appropriate to the observation phase (Epoch 3),
and show it in figure~\ref{fig:PIN_qui} (b).
We also summarized the statistics of the data, the NXB, and the CXB in table~\ref{table:HXD_log_MWLII}.
The NXB count rate for on-source is higher than that for the Earth occultation,
due to different orbit conditions of the satellite, as can be seen in
tables~\ref{table:HXD_log_MWLI} and \ref{table:HXD_log_MWLII}.
Therefore, it is of no use to perform a direct comparison of the count rates between the Earth
occultation and on-source periods, 
and therefore we examined the ratio of the combined NXB and CXB to the NXB.    
After subtracting the CXB and NXB models, 
the on-source data corresponded to 
4.2 $\pm$ 0.3 $\%$ and $-0.9 \pm 0.2$ $\%$ of the NXB, 
in 12 -- 20 keV and 20 -- 40 keV, respectively.
By using the Earth occultation data, we ascribed the apparent excess in 12 -- 20 keV 
due to the uncertainly of the NXB model instead of the source signals
since we confirmed that the NXB model underestimates the
real NXB spectrum by $\sim 4$~\%.  

Next, in (a) in figure~\ref{fig:PIN_flare}, we show the HXD-PIN spectra 
of MWL~II during the Earth occultation with an exposure of 28.7 ks.
Table~\ref{table:HXD_log_MWLII} summarizes the statistics of the data
and the NXB model during the Earth occultation.
They are consistent with each other within $1.4 \pm 1.2 $\% 
and $1.2 \pm 1.6 $\%  in 12 -- 20 keV and 20 -- 40 keV. 
Thus, we confirmed that the NXB model adequately reproduced the observed
NXB spectrum in MWL~II within its reported uncertainty.

Figure~\ref{fig:PIN_flare} (b) shows the on-source spectra in MWL~II.
In order to fold the CXB spectra,  
we used the HXD-PIN response function (Epoch~4), 
{\tt ae\_hxd\_pinhxnome4\_20080129.rsp}.
The HXD-PIN spectrum revealed a significant excess 
over the combined CXB and NXB in the $12$ -- $27$~keV range. 
The count rates for the data, the NXB and the CXB in this range, are shown in table~\ref{table:HXD_log_MWLII}.
Thus, the excess has a count rate of 
$0.011 \pm 0.002$ cts s$^{-1}$ ($2.9 \pm 0.6$ \% of the NXB),
which corresponds to statistical significance of $ 5.0 \sigma$.
Since this is well above the NXB errors,
we concluded that the HXD-PIN detected signals 
from OJ~287 in MWL~II. 

Due to the rather low signal statistics of the HXD-PIN in comparison to the XIS, 
instead of joint fitting of XIS and HXD-PIN spectra, we compared the flux of the HXD-PIN signal to the fitting result of the
XIS spectrum.  
The simple extrapolation of the best-fit single PL model to the XIS
spectrum (section~\ref{sec:XIS})
has a count rate of $0.016$ cts s$^{-1}$ in 12 -- 27 keV,
after the correction of the relative normalization of the HXD to the XIS 
(1.14: \cite{XIS2HXD}). 
The detected HXD-PIN count rate is only 0.65 times this value, which
suggests the presence of a spectral break between the XIS and the HXD-PIN range, 
although it is statistically insignificant (2.6~$\sigma$).

\section{Multi-wavelength Observations and Results}
\subsection{Radio Observations}
Eleven radio observations of OJ~287 with the NMA 
at the Nobeyama Radio Observatory were carried out 
between January 2007 and January 2008. 
The NMA consists of six 10-m antennas equipped with cooled DSB SIS receivers. 
An Ultra-Wide-Band Correlator~\citep{Okumura} was employed as the spectrocorrelator. 
We performed simultaneous observations of a radio continuum emission from OJ~287 at 86.75~GHz and 98.75~GHz, with a 1~GHz bandwidth for each band.
The total on-source time of OJ~287 in each observation 
was between nine and twelve minutes. 
In each observation session, 
OJ~287 and the bright reference amplitude calibrators (3C~84 or 3C~345) were observed alternately. 
The relative amplitude ratios between OJ~287 and these reference calibrators were obtained from the visibility data.
As absolute flux-scale calibrations, Uranus or Neptune was used in order to calculate the final flux values of OJ~287.
The uncertainties in the absolute flux scale were estimated to be about 15~\% 
for each observation, taking into account statistical errors 
induced by noise and systematic errors caused by flux fluctuations of the reference calibrators.

For MWL~I, we observed the source on $11$ and $12$ April, overlapping with the Suzaku pointing.
The averaged flux of the two night observations was 
$1.73 \pm 0.26$~Jy and $1.75 \pm 0.26$~Jy at $86.75$~GHz and $98.75$~GHz, respectively. 
During the period of MWL~II, the observations on November $7$ and $8$ were coincident with the Suzaku pointing.
The averaged flux of the two day observations was 
$3.04 \pm 0.46$~Jy and $2.98 \pm 0.45$~Jy at $86.75$~GHz and $98.75$~GHz, 
respectively (see table~\ref{table:radio_opt_log}).    
The measured radio flux for MWL~II was $1.7$ -- $1.8$ times higher than that for MWL~I.

\subsection {Optical Observations}
\label{sec:optical_obs}
Optical and near-infrared (NIR) photometric observations were performed 
at the Higashi-Hiroshima Observatory with the $1.5$-m ``KANATA'' telescope. 
We obtained images of the field of OJ~287 with $V$, $J$, 
and $K_s$ filters by using TRISPEC attached to the telescope 
\citep{Watanabe}.  
The $V$-band images were observed for $53$ nights between $2006$ October 
and $2007$ December. 
For each night, we obtained $\sim 20$ images with an exposure time of $108$~s for each frame.  
The magnitude of OJ~287 was measured by differential photometry with a neighbor comparison star 
located at 
($\alpha, \delta$) = (\timeform{08h54m52.7s}, \timeform{20D04'46''}).
The $V$-magnitude of the comparison star ($V=14.160$) was quoted from
\citet{Skiff}. After preparing dark-subtracted and flat-fielded images,  
we measured the magnitudes of the objects by using the aperture photometry 
package in IRAF.
We also checked the constancy of the comparison star by using a neighbor 
star at 
($\alpha, \delta$) = (\timeform{08h54m59.0s}, \timeform{20D02'58''}).
The differential magnitude of the comparison star exhibited no significant
fluctuation exceeding $0.003$-th during our observations.

We also obtained NIR images, 
for MWL~I on 2007 April $11$, $12$, and $13$, as well as 
for MWL~II on November $7$, $8$, and $10$, which overlapped with the Suzaku pointing.   
Simultaneous $V$, $J$, and $K_s$-band images are available 
for these six nights. 
The exposure times of each frame were $15$ and
$4$~s for the $J$ and $K_s$-band images, respectively.  
The reduction procedure was similar to 
that in the $V$-band data mentioned above, and we used the same comparison stars and 
target stars as those for the $V$-band data.  
The $J$ and $K_s$ magnitudes of the comparison star were quoted from 
the 2MASS catalog ($J=12.664$ and $K_s=12.090$; \cite{Skrutskie}).  
We confirmed that the magnitude of the comparison star remained constant 
within $0.004$ and $0.012$-th for $J$ and $K_s$-bands during our observations.
In MWL~I, 
the averaged flux of three night observations was 
$3.03\pm 0.01$~mJy, $8.81 \pm 0.03$~mJy, and $17.74 \pm 0.33$~mJy 
for $V$, $J$, $K_{s}$-band, respectively.
In MWL~II, these fluxes were
$8.93 \pm 0.05$~mJy, $27.02 \pm 0.21$~mJy, and $55.95^{+7.96}_{-6.76}$~mJy, respectively. 
(see table~\ref{table:radio_opt_log}).
It can be seen that the source flux for MWL~II is higher by a factor $3$ as compared
to that for MWL~I.
The optical data show a monotonous decrease in a time scale of $\sim$ 4
days during MWL~II by a factor of 1.3.

In addition, we obtained $R$-band frames from the KVA Telescope simultaneously with Suzaku pointings. 
The exposure time was 180~s for each frame. 
Photometric measurements were conducted in differential mode, 
that is, by obtaining CCD images of the target and the calibrated comparison stars in the same field of view (\cite{F_and_T}). 
The magnitudes of the source and the comparison stars were measured 
using aperture photometry and the (color-corrected) zero point of the image 
as determined from the magnitude of the comparison star. 
Finally, the object magnitude was computed using the zero point and a filter-dependent color correction. 
After this, magnitudes were transferred into linear fluxes by using the formula 
$F = F_{0} \times 10^{(mag/-2.5)}$,
where mag is the magnitude of the object and $F_{0}$ is a filter-dependent zero point 
(the value $F_{0}$ = 3080~Jy is used in the $R$-band, \cite{Bessell}).

\subsection{VHE $\gamma$-ray Observations}
We used the MAGIC telescope to search for VHE $\gamma$-rays 
emission from OJ~287 during the MWL campaigns I and II.
MAGIC is 
a single dish Imaging Atmospheric Cherenkov Telescope (IACT) with a 17-m diameter main reflector. 
The telescope is located in the Canary Island of La Palma, in regular operation since 2004
with a low energy threshold of $50$ -- $60$ GeV 
(trigger threshold at small zenith angles;~\cite{MAGICCrab}). 

In MWL~I, MAGIC observed during $3$ nights. 
The zenith angle of the observations ranges from $8^{\circ}$ to
$29^{\circ}$. The observations were performed in so-called ON-OFF
observation mode. The telescope was pointing directly to the source,
recording ON-data. The background was estimated from
additional observations of regions where no $\gamma$-ray is expected,
OFF-data, which were taken with sky conditions similar to ON-data. 
Data runs with anomalous trigger rates due to bad observation conditions
were rejected from the analysis. 
The remaining data correspond to $4.5$ hours of ON and $6.5$ hours of OFF data.
In November and December $2007$ for MWL~II,
MAGIC observed
in a zenith angle range from $8^{\circ}$ to $31^{\circ}$ in the 
``wobble mode''~\citep{wobble}, where the object was observed at
$0.4^{\circ}$ offset from the camera center. 
With this observation mode,
an ON-data sample and OFF-data samples can be extracted from the same observation run; 
in our case, we used 3 OFF regions to estimate the background.
In total, the data were taken during $22$ 
nights, two of which coincide  
with the Suzaku pointing. 
$41.2$ hours of data from $19$ nights passed the quality selection to be
used for further analysis. 

The VHE $\gamma$-ray data taken for MWLs~I and II were analyzed using the MAGIC
standard calibration and analysis software. 
Detailed information about the analysis chain is found
in~\citet{MAGICCrab}.  
In February $2007$, the signal digitization of MAGIC was upgraded
to $2$ $\rm{GSamples~s^{-1}}$ FADCs, and timing information is used to
suppress the 
contamination of light of the night sky and to obtain new shower image
parameters~\citep{MAGICtime} in addition to conventional \textit{Hillas} image
parameters~\citep{Hillas}. 

These parameters were used for $\gamma$/hadron separation by means of
the ``Random Forest (RF)'' method~\citep{MAGICRF}. 
The $\gamma$/hadron separation based on the RF method was turned to give a $\gamma$-cut efficiency of $70~\%$. 
Finally, the $\gamma$-ray signal was determined by comparing between ON and normalized-OFF data 
in the |ALPHA| parameter\footnote{the angle between the shower image principal axis
and the line connecting the image center of gravity with the camera center.} 
distribution, in which the $\gamma$-ray signal should show up as an excess at small values. 
Our analysis requires a $\gamma$-cut efficiency of $80~\%$ for the final |ALPHA| selection.
The energy of the $\gamma$-ray events are also estimated using the RF method.

A search of VHE $\gamma$-rays from OJ~287 was performed with data
taken for MWLs~I and II in three distinct energy bins.
No significant excess was found in any data samples. 
Upper limits with $95~\%$ confidence level in the number of excess events
were calculated using the method of \citet{Rolke}, taking into
account a systematic error of 30~\%. The number of excess events was converted
into flux upper limits assuming a photon index of $- 2.6$, 
corresponding to the value used in our Monte-Carlo samples of $\gamma$-rays.
The derived upper limits in the three energy bins for each period are summarized in table~\ref{table:UL_log}. 

\begin{figure}[ht]
\begin{center}
\FigureFile(80mm,80mm){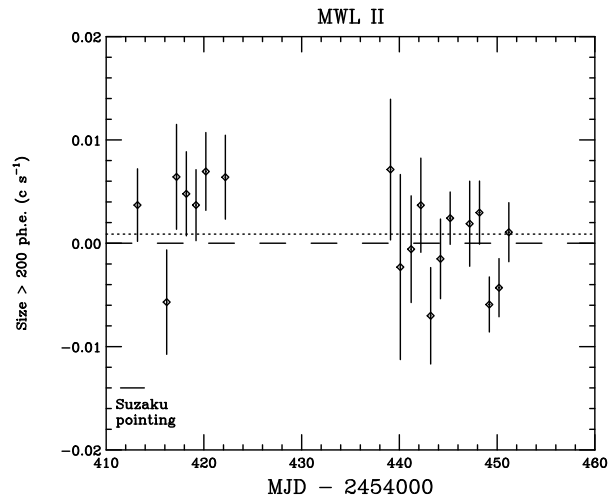}
\end{center}
\caption{
Excess event rate with SIZE above 200 photoelectrons 
(with a corresponding energy threshold of 150~GeV), 
observed with the MAGIC telescope in MWL~II.
The dotted line indicates the average count rate. 
}
\label{fig:magic_LC}
\end{figure}

A search for VHE flares with a short-time scale was also performed with the data taken for MWL~II. 
Figure~\ref{fig:magic_LC} shows the nightly count rate of the excess
events after all cuts including a SIZE cut above 200 photoelectrons,
corresponding to an energy threshold of 150 GeV. Fitting 
a constant rate to the observed flux yields 
$\chi^2/{\rm d.o.f.} = 25.55/18$ (a probability of $11~\%$), and thus indicates 
no evidence of a VHE flare during
this period.

\section{Discussion}
\begin{figure*}[ht]
\begin{center}
\FigureFile(100mm,100mm){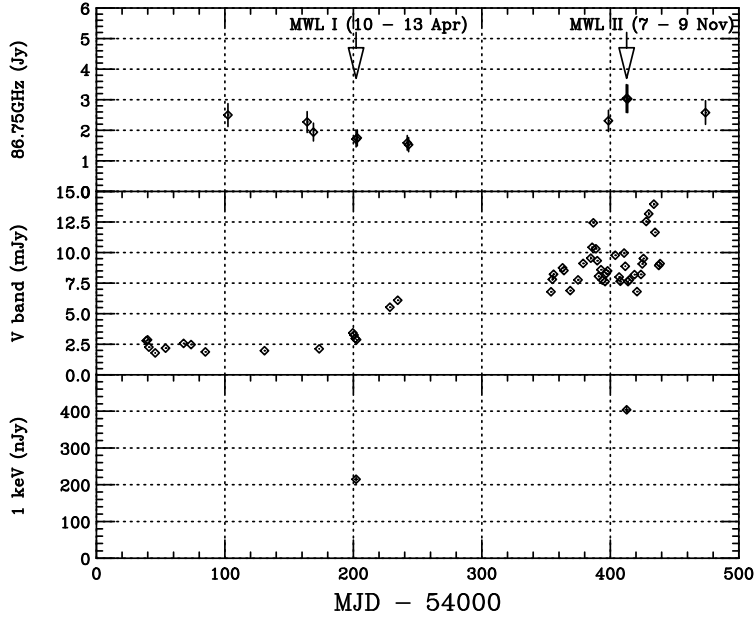}
\end{center}
\caption{
The multi-wavelength lightcurves of OJ~287.
The top panel shows the radio flux at $86.75$~GHz as observed with NMA, while
the middle panel shows the optical flux in the $V$-band as observed with KANATA.
The radio and optical fluxes are averaged over each night.
The bottom panel shows the X-ray flux density at 1~keV.
Arrows indicate the Suzaku pointings in MWL~I and MWL~II.
}  
\label{fig:multi_lc}
\end{figure*}

We performed Suzaku X-ray observations of OJ~287 in the quiescent state
(MWL~I) and the second flare (MWL~II) in $2007$ April and November,
respectively, where 
the latter was the first X-ray observation during the second flare of
the source.
In cooperation with Suzaku, 
radio, optical, and VHE $\gamma$-ray observations were
performed with NMA, KANATA, and MAGIC, respectively.
Figure~\ref{fig:multi_lc} summarizes the multi-wavelength lightcurves obtained 
between September 2006 and January 2008.
While the optical flux of OJ~287 was below $3~\rm{mJy}$ in 
the $V$-band before MWL~I,  
the brightness of the source started increasing after MWL~I to become the flaring
state ($\gtrsim 7$ mJy) in September $2007$.
Although the radio flux gradually decreased around the period of MWL~I,  
it started increasing during the optical flare in MWL~II, 
and it also increased in 
other energy bands.  
We found that the brightness of the source increased by a factor of $2$ -- $3$, 
in the radio, optical, and X-ray bands between MWL~I and MWL~II,
although no significant VHE $\gamma$-ray signals were detected in either MWL~I or MWL~II.

As shown in figure \ref{fig:XIS_spec},
significant X-ray signals were detected with the XIS in the range $0.5$ -- $10$ keV 
range. 
The HXD-PIN detected hard X-ray signals in $12$ -- $27$~keV 
with a significance of $5.0~\sigma$ in MWL~II 
(figure \ref{fig:PIN_flare}), 
while those signals were not significant in MWL~I (figure \ref{fig:PIN_qui}). 
The XIS spectra in $0.5$ -- $10$~keV were described with a single
PL model modified with Galactic absorption in MWL~I and MWL~II.
The photon indices and the flux densities at $1$ keV were derived as 
$1.65 \pm 0.02$ and $215 \pm 5$~nJy in MWL~I,
and $1.50 \pm 0.01$ and $404^{+6}_{-5}$~nJy in MWL~II. 

In the previous observations of the first flares and quiescent states,
the X-ray flux densities at 1~keV and photon indices of OJ~287 were
estimated as 
0.15 -- 0.31 $\mu$Jy          and 1.45 -- 1.63, respectively, in 1997 -- 2001 with BeppoSAX \citep{BeppoSAX},   
0.22 -- 0.25 $\mu$Jy          and 1.51 -- 1.57    in 1997      with ASCA \citep{Isobe},
0.76 $^{+0.03}_{-0.06}~\mu$Jy and 1.67 $\pm$ 0.02 in 1994      with ASCA \citep{Idesawa},
2.08 -- 2.24 $\mu$Jy          and 2.16 -- 2.37 in 1983 -- 1984 with EXOSAT \citep{EXOSAT}, and
0.94 -- 2.70 $\mu$Jy          and 1.5  -- 2.3  in 1979 -- 1980 with Einstein satellite \citep{Einstein}. 
These results suggest that the X-ray spectrum became softer
as the source became brighter (\cite{Idesawa,Isobe}).
This trend was interpreted in the following way: 
during the quiescent state, the IC component dominates the X-ray spectrum \citep{Isobe}.
However, once the X-ray flux increases, the high-frequency end of the SR component
extends to the X-ray band and exhibits a softer spectrum.
In fact, \citet{Isobe} have successfully decomposed 
the X-ray spectrum obtained with ASCA at the first flare in $1994$ 
into soft and hard PL components
representing the SR and the IC components, respectively. 
Moreover, in the XMM-Newton observation of the recent first flare in $2005$, 
the source exhibited a concave broken-PL-like X-ray spectrum (\cite{Ciprini})
which also supports the idea that there is a contribution from the soft component.
On the other hand, the spectral behavior
in the second flare of OJ~287 observed with Suzaku 
was completely the opposite to the previous X-ray trend in the first flares.

In order to perform a quantitative evaluation of the possible soft excess component
in the Suzaku X-ray spectrum obtained in MWL~II,  
we employed an additional steep PL model modified with the Galactic absorption. 
The photon index of the additional steep PL component was fixed
at $2.62$, which is the best fit value for the first flare in $1994$ (\cite{Isobe}),  
while its normalization was left free.
As a result of this two-component model fitting,
the upper limit on the flux density of the additional soft PL component 
was derived as $8.7$~nJy at the $3 \sigma$ level (see table~\ref{table:XIS_log}),
while the hard component remained consistent with the best fit values for the single PL model fitting ($\chi^2/{\rm d.o.f.} = 566.4/519$).
Thus, we obtained the upper limit on the ratio of the soft to the hard components at 1~keV
as $0.022$,
although the ratio was estimated to be $0.186 \pm 0.034$
in the first flare of $1994$ (\cite{Isobe}). 
Therefore, the contribution from the soft component in MWL~II was 
negligible in comparison to that of the first flare, 
and the hard component fully dominated the XIS spectrum 
over the 0.5 -- 10 keV range.

\begin{figure}[ht]
\begin{center}
\FigureFile(100mm,100mm){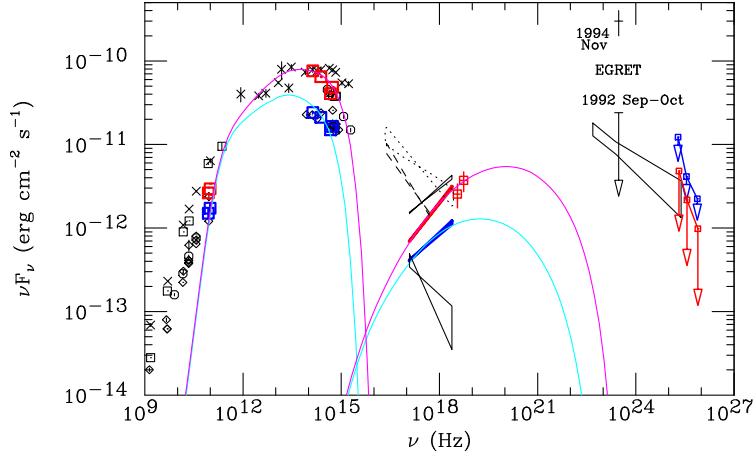}
\end{center}
\caption{
The SED of OJ~287 during the Suzaku observations in MWL~I (blue) and MWL~II (red). 
The radio and optical data are shown with squares and,
the X-ray data are shown with bow ties.  
The upper limit of the VHE $\gamma$-ray spectrum are measured values, 
shown with downward arrows.
The light blue lines and the purple lines indicate 
the simple one-zone SSC model for MWL~I and MWL~II, respectively. 
The black data points show radio, optical, and $\gamma$-ray data 
from non-simultaneous observations. 
The X-ray spectra with EXOSAT, ROSAT, and ASCA are 
drawn with dotted, dashed, and solid lines, respectively 
(\cite{Idesawa,Isobe} and references therein).
The $\gamma$-ray spectrum obtained with Fermi during the first 3 month 
observation (August -- October 2008) is shown with a bow tie~\citep{fermi_detect}. 
}
\label{fig:SED}
\end{figure}

Figure~\ref{fig:SED} shows the overall SED of OJ~287 
for the Suzaku pointing in MWL~I and MWL~II,
including the optical $R$-band data from the KVA Telescope,
as well as some historical data.
In the figure, we recognize two spectral components, 
which are typical of blazars.
The low frequency SR component, 
extending from radio to optical frequencies,
has a spectral turnover at around $5 \times 10^{14}$ Hz.
At the same time,
the observed SR component is well above 
the extrapolation from the upper limit of the soft PL component in MWL~II 
($\sim 1 \times 10^{-12} {\rm erg~s^{-1}~cm^{-2}}$ at $5 \times 10^{14}$ Hz).
Therefore, we naturally attribute the observed hard X-ray spectrum 
to the IC component rising toward the higher frequency range.
The SED indicates that 
both the SR and IC intensities increased from MWL~I to MWL~II
without any significant shift of the SR peak frequency.

As a working hypothesis, here we assume simply that
the variation 
of the SED was caused by a change 
in electron energy density (or number density) 
and/or the maximum Lorentz factor of the electrons,
with stable magnetic field, volume of emission region, 
minimum Lorentz factor, and break of electron energy distribution (e.g., \cite{Mrk421}).
In order to evaluate this hypothesis, 
we applied a one-zone SSC model to the SED
by using the numerical code developed by \citet{Kataoka}.  
The electron number density spectrum was assumed to be a broken PL and  
the index of the electron spectrum ($p$) below the break Lorentz factor was
determined by the X-ray photon index as $p=2\Gamma - 1 = 2.3$ and $2.0$, 
in MWL~I and MWL~II, respectively.  
We obtained the following seven free parameters to describe the observed SED: 
the Doppler factor ($\delta$), the electron energy density ($u_{\rm e}$), 
the magnetic field ($B$), the blob radius ($R$), 
and the minimum, break, and maximum Lorentz factor of the electrons 
($\gamma_{\rm min}$, $\gamma_{\rm break}$, and $\gamma_{\rm max}$, respectively).
Adopting the optical variability time scale ($T_{\rm var} \sim$ 4 days; section~\ref{sec:optical_obs}) in MWL~II, 
the relation between $\delta$ and $R$ should be subjected to 
$R < c T_{\rm var} \delta / (1+z) = 1.2 \times 10^{17} (T_{\rm var} / 4{\rm days}) (\delta / 15)~{\rm cm}$
where $c$ and $z$ are the speed of light and the redshift of the source,
respectively.

We derived the SSC model parameters 
as summarized in table~\ref{table:SSC_log}.
The resultant model curves are shown 
with solid lines in figure~\ref{fig:SED}.
The both model predictions are well below the upper limit on the VHE $\gamma$-ray spectra. 
The SED in MWL~I was reproduced with 
$\delta=15$, 
$B=0.71$~G, 
$R=7.0 \times 10^{16}~\rm{cm}$, 
$\gamma_{\rm {min}}=70$, 
$\gamma_{\rm break}~=~700$, 
$\gamma_{\rm max}=3300$, and 
$u_{\rm e}=1.5 \times 10^{-3}~\rm{erg~cm^{-3}}$. 
The $\delta$ and $R$ are typical values for LBL \citep{Ghi98}.
On the other hand, in the SED in MWL~II, 
the SSC model with 
$u_{\rm e}=2.1 \times 10^{-3}~\rm{erg~cm^{-3}}$ and
$\gamma_{\rm {max}}=4500$ 
was found to describe the SED successfully, 
while the other parameters remained unchanged.
Thus, we adopt the interpretation that 
the increase in the electron energy density produced the second flare.

The SED spectra obtained in 1st and 2nd flare during 2005 -- 2008
outburst are suggested to have different features as we saw in the X-ray
spectra obtained with XMM-Newton and Suzaku.
The difference in these SED spectra may require not only simple `disk
impact' \citep{Valtonen_nature},
but also a state transition of the disk - jet system \citep{disk-jet}.

In 2008, Fermi successfully detected a $\gamma$-ray spectrum from the quiescent state of OJ~287 during
its first three months \citep{fermi_detect}, as shown in figure~\ref{fig:SED}.
The measured $\gamma$-ray flux significantly exceeds our simple SSC model flux.
This may indicate a contribution of external Compton (EC) radiation to the $\gamma$-ray emission from OJ~287. 
Assuming that the $\gamma$-ray spectrum peaks around $\sim 1\times 10^{22}$ Hz
with a flux of $\sim 2 \times 10^{-11}$ erg~s$^{-1}$ cm$^{-2}$,
as is suggested from figure~\ref{fig:SED},
and using the electron spectrum determined
from the SSC fitting to the radio-to-X-ray spectrum derived for MWL~II,
the energy density and typical frequency of
$u_{\rm seed} \sim 10^{-5}$ erg~cm$^{-3}$ and $\nu_{\rm seed}\sim 10^{14}$ Hz
in the rest frame of the nucleus (not the jet frame)
are required for the seed photons of the EC process.
These give the luminosity of the seed photon source
as $\sim 10^{42}$ erg~s$^{-1}$.
This luminosity is consistent with that evaluated for 3C~279 \citep{Inoue},
a famous quasar-hosted blazar of which the $\gamma$-ray spectrum is thought
to be dominated by the EC emission, within an order of magnitude.
However, for a precise identification of the seed photon source,
it is crucial to make a simultaneous multi-wavelength observation
from radio to $\gamma$-ray frequencies. 

\section{Summary}
We performed X-ray observations of OJ~287 
in the quiescent state (MWL~I) in 2007 April and 
in the flaring state (MWL~II) in 2007 November by using Suzaku, 
together with radio, optical, and VHE $\gamma$-ray observations 
with NMA, KANATA, and MAGIC, respectively.
The obtained results can be summarized as follows.
\begin{itemize}
\item The brightness of OJ~287 increased by a factor of 2 
      from MWL~I with a 1~keV flux density of $215 \pm 5$ nJy 
      to MWL~II with $404^{+6}_{-5}$ nJy.
      The X-ray spectrum of OJ~287 was harder in MWL~II (a photon index
      of $\Gamma = 1.50 \pm 0.01$) than in MWL~I ($\Gamma = 1.65 \pm 0.02$).
\item In MWL~II, hard X-ray signals from the object were detected with a significance of
      $5.0 \sigma$ in the 12 -- 27 keV range,
      for the first time.
\item The radio and optical fluxes doubled from MWL~I to MWL~II.
      No significant VHE $\gamma$-ray signals were detected in either observation.
\item In both observations, the object exhibited a typical blazar-like spectral energy distribution
      consisting of synchrotron and inverse Compton components.
      The relative hard X-ray spectrum appeared to be dominated by the inverse Compton
      components.
\item No significant difference of the synchrotron cut-off or peak frequency
      was found between the quiescent and flaring states.
      This spectral behavior is rather different from that in past flares,
      in which it appeared that the synchrotron component  contributed to the X-ray
      spectrum.
\item Based on the synchrotron self-Compton model, 
      the change 
      of the multi-wavelength spectral energy distribution is interpreted as
      the increase of the electron energy density,
      without any notable change in either the magnetic field 
      or the electron Lorentz factor. 
\end{itemize}

\bigskip

We thank all members of the Suzaku team
for performing successful operatio and calibration.
The Nobeyama Radio Observatory is a branch of the National Astronomical
Observatory of Japan, the National Institutes of Natural Sciences (NINS).
IRAF is distributed by the National Optical 
Astronomy Observatories, which are operated by the Association of 
Universities for Research in Astronomy, Inc., under a cooperative 
agreement with the National Science Foundation.
We would like to thank the Instituto de Astrofisica de
Canarias for the excellent working conditions at the
Observatorio del Roque de los Muchachos in La Palma. 
The support of the German BMBF and MPG, the Italian INFN 
and Spanish MICINN is gratefully acknowledged. 
This work was also supported by ETH Research Grant 
TH 34/043, by the Polish MNiSzW Grant N N203 390834, 
and by the YIP of the Helmholtz Gemeinschaft.
Dr. J. Kataoka kindly provided valuable information
on the numerical calculation for the wide-band spectrum of blazars.
N. I. is supported by the Grant-in-Aid for the Global COE Program,
"The Next Generation of Physics, Spun from Universality and Emergence"
from the Ministry of Education, Culture, Sports,
Science and Technology (MEXT) of Japan.


\begin{longtable}[ht]{lcccccccc}
\caption{Summary of model fitting to the Suzaku XIS spectra.}
\label{table:XIS_log}
\hline \hline
  obs & Models & $N_{\rm H}$\footnotemark[$*$] & 
 $\Gamma$                          & $S_{\rm 1keV}$\footnotemark[$\dagger$] & 
 $\Gamma$\footnotemark[$\ddagger$] & $S_{\rm 1keV}$\footnotemark[$\dagger$]\footnotemark[$\ddagger$] &
 $\chi^2 / {\rm d.o.f.}$ & $F_{2 - 10 {\rm keV}}$\footnotemark[$\S$]\\ 
\hline 
\endfirsthead
\hline\hline
 & & & & &\\ 
\hline
\endhead
\hline
\endfoot
\hline
\multicolumn{5}{@{}l@{}}{\hbox to 0pt{\parbox{100mm}{\footnotesize
\footnotemark[$*$] in $10^{20}$ cm$^{-2}$.
\par\noindent
\footnotemark[$\dagger$] X-ray flux density at 1 keV, in units of nJy. 
\par\noindent
\footnotemark[$\ddagger$] Soft PL component parameter.
\par\noindent
\footnotemark[$\S$] 2 -- 10 keV flux in ${\rm erg~cm^{-2}~s^{-1}}$.
\par\noindent
\footnotemark[$\|$] Fixed at the Galactic Value. 
}\hss}}
\endlastfoot
MWL~I  & PL        & $2.56$\footnotemark[$\|$] & $1.65~\pm~0.02$        & $215~\pm~5$     & --             & --      & $215.0/243$ & $1.44 \times 10^{-12}$\\ 
MWL~II & PL        & $2.56$\footnotemark[$\|$] & $1.50~\pm~0.01$        & $404^{+6}_{-5}$ & --             & --      & $566.3/520$ & $3.42 \times 10^{-12}$\\ 
       & double PL & $2.56$\footnotemark[$\|$] & $1.50^{+0.01}_{-0.02}$ & $404^{+4}_{-9}$ & $2.62$ (Fixed) & $<~8.7$ & $566.4/519$ & $3.42 \times 10^{-12}$\\ 
\end{longtable}

\begin{longtable}[ht]{lcccc}
\caption{Event statistics of the HXD data in MWL~I.}
\label{table:HXD_log_MWLI}
\hline \hline
\multicolumn{1}{l}{} & \multicolumn{2}{c}{Earth occulation} & \multicolumn{2}{c}{On-source} \\
   & 12 -- 20 keV         & 20 -- 40 keV         & 12 -- 20 keV          & 20 -- 40 keV  \\ 
\hline 
\endfirsthead
\hline\hline
 & & & &  \\ 
\hline
\endhead
\hline
\endfoot
\hline
\multicolumn{5}{@{}l@{}}{\hbox to 0pt{\parbox{140mm}{\footnotesize
\footnotemark[$*$] Data excess over the NXB and the combined NXB and CXB for Earth occultation and on-source, respectively.
\par\noindent
\footnotemark[$\dagger$] Ratio to the NXB model.
}\hss}}
\endlastfoot
Data (c~s$^{-1}$)                     & 0.274 $\pm$ 0.003 & 0.149 $\pm$ 0.002 & 0.319 $\pm$ 0.002 & 0.173 $\pm$ 0.001\\
NXB  (c~s$^{-1}$)                     & 0.264 $\pm$ 0.001 & 0.150 $\pm$ 0.001 & 0.290 $\pm$ 0.001 & 0.167 $\pm$ 0.001 \\
CXB  (c~s$^{-1}$)                     & --                & --                & 0.016             & 0.008\\
excess (c~s$^{-1}$)\footnotemark[$*$] & 0.010 $\pm$ 0.003 & $-$0.015$\pm$0.002& 0.012 $\pm$ 0.002 & $-0.002 \pm$ 0.001\\
excess\footnotemark[$*$] ratio\footnotemark[$\dagger$] (\%)
                                      & 3.8   $\pm$ 1.0   & $-$1.0 $\pm$ 1.3  & 4.2 $\pm$ 0.3     & $-0.9\pm$ 0.2\\
\end{longtable}

\begin{longtable}[ht]{lccc}
\caption{The event statistics of the HXD data in MWL~II.}
\label{table:HXD_log_MWLII}
\hline \hline
\multicolumn{1}{l}{} & \multicolumn{2}{c}{Earth occultation} & \multicolumn{1}{c}{On-source} \\
 & 12 -- 20 keV         & 20 -- 40 keV         & 12 -- 27 keV          \\   
\hline 
\endfirsthead
\hline\hline
 & & &  \\ 
\hline
\endhead
\hline
\endfoot
\hline
\multicolumn{5}{@{}l@{}}{\hbox to 0pt{\parbox{140mm}{\footnotesize
\footnotemark[$*$] Data excess over the NXB and the combined NXB and CXB for Earth occultation and on-source, respectively.
\par\noindent
\footnotemark[$\dagger$] Ratio to the NXB model.
}\hss}}
\endlastfoot
Data (c~s$^{-1}$)                     & 0.286 $\pm$ 0.003   & 0.165 $\pm$ 0.002    & 0.399 $\pm$ 0.002  \\
NXB  (c~s$^{-1}$)                     & 0.290 $\pm$ 0.001   & 0.167 $\pm$ 0.001    & 0.367 $\pm$ 0.001  \\
CXB  (c~s$^{-1}$)                     & --                  & --                   & 0.022             \\
excess (c~s$^{-1}$)\footnotemark[$*$] &$-$0.004 $\pm$ 0.003 & $-$0.002 $\pm$ 0.003 & 0.011 $\pm$ 0.002 \\
excess\footnotemark[$*$] ratio\footnotemark[$\dagger$]
                                      & $-1.4 \pm$ 1.2      & $-1.2 \pm$ 1.6        &  2.9 $\pm$ 0.6\\
\end{longtable}

\begin{longtable}[ht]{lcccccc}
\caption{Summary of radio and optical fluxes obtained during the Suzaku pointing in MWL~I and MWL~II.}
\label{table:radio_opt_log}
\hline \hline  
  obs & \multicolumn{2}{c}{radio flux (Jy)}  & \multicolumn{4}{c}{optical flux (mJy)} \\ 
      & 86.95~GHz\footnotemark[$*$] & 98.75~GHz\footnotemark[$*$] & $K_{s}$ \footnotemark[$\dagger$]& $J$ \footnotemark[$\dagger$]& $R$ \footnotemark[$\ddagger$]& $V$\footnotemark[$\dagger$] \\
\hline 
\endfirsthead
\hline\hline
 & & & & & \\ 
\hline
\endhead
\hline
\endfoot
\hline
\multicolumn{7}{@{}l@{}}{\hbox to 0pt{\parbox{140mm}{\footnotesize
\footnotemark[$*$] NMA data.
\par\noindent
\footnotemark[$\dagger$] KANATA data.
\par\noindent
\footnotemark[$\ddagger$] KVA Telescope data.
}\hss}}
\endlastfoot
MWL~I  & $1.73 \pm 0.26$ & $1.75 \pm 0.26$ & $17.74 \pm 0.33$        & $8.82 \pm 0.03$  & $3.20 \pm 0.05$ & $3.03 \pm 0.01$ \\
MWL~II & $3.04 \pm 0.46$ & $2.98 \pm 0.46$ & $55.95^{+7.69}_{-6.76}$ & $27.02 \pm 0.21$ & $8.70\pm0.14$   & $8.93 \pm 0.05$ \\
\end{longtable}

\begin{longtable}[ht]{lccc}
\caption{Results of the search for VHE $\gamma$-ray emissions from OJ~287. }
\label{table:UL_log}     
\hline \hline
MWL~I & & &\\
\hline 
\endfirsthead
\hline\hline
 & & & \\ 
\hline
\endhead
\hline
\endfoot
\hline
\multicolumn{4}{@{}l@{}}{\hbox to 0pt{\parbox{125mm}{\footnotesize
\footnotemark[$*$] Correspond to peak energies of $\gamma$-ray Monte Carlo samples after all cuts.
\par\noindent
\footnotemark[$\dagger$] Number of measured ON events.
\par\noindent
\footnotemark[$\ddagger$] Normalized number of OFF events and related error.
\par\noindent
\footnotemark[$\S$] Based on equation~(17) in~\citet{LiMa},
\par\noindent
\footnotemark[$\|$] 95\% upper limit of the number of excess events with 30\% systematic error.
\par\noindent
\footnotemark[$\#$] Flux upper limit assuming a photon index of $-2.6$ for the calculation of the effective area.
\par\noindent
\footnotemark[$**$] Corresponding Crab flux in each energy range based on measurements of the Crab pulsar performed with the MAGIC telescope ~\citep{MAGICCrab}
}\hss}}
\endlastfoot
Threshed Energy (GeV)\footnotemark[$*$]    & $80$          & $145$       & $310$\\
ON events\footnotemark[$\dagger$]          & $40056$       & $1219$      & $42$\\
OFF events\footnotemark[$\ddagger$]        & $40397\pm226$ & $1340\pm38$ & $39.5\pm6.3$\\
significance ($\sigma$)\footnotemark[$\S$] & $-1.13$       & $-0.94$     & $-0.47$\\
U.L. of excess events\footnotemark[$\|$]   & $394$         & $75.1$      & $21.9$\\
Flux$_{95\% \rm{U.L.}}$ ($\times10^{-12}$ cm$^{-2}$ s$^{-1}$)\footnotemark[$\#$]  
                                           & $59.8$        & $11.1$      & $2.83$\\
Crab Flux(\%)\footnotemark[$**$]           &   $8.5$       &  $3.3$      &  $2.4$ \\
\hline  \hline 
MWL~II  & & &\\
\hline 
Threshed Energy (GeV)\footnotemark[$*$]    & $85$          & $150$       & $325$\\
ON events\footnotemark[$\dagger$]          & $281885$      & $12582$     & $578$\\
OFF events\footnotemark[$\ddagger$]        & $282342\pm493$& $12573\pm65$& $576\pm14$ \\
significance ($\sigma$)\footnotemark[$\S$] & $-0.75$       & $0.07$      & $0.07$\\
U.L. of excess events\footnotemark[$\|$]   & $1218$        & $330$       & $71.6$\\
Flux$_{95\% \rm{U.L.}}$ ($\times10^{-12}$ cm$^{-2}$ s$^{-1}$)\footnotemark[$\#$]
                                           & $22.1$        & $5.64$      & $1.18$\\
Crab Flux(\%)\footnotemark[$**$]           &  $3.4$        &     $1.7$   &  $1.1$\\
\end{longtable}

\begin{longtable}[ht]{lcc}
\caption{Physical parameters for the SSC model.}
\label{table:SSC_log}
\hline \hline 
parameters\footnotemark[$*$] & MWL~I & MWL~II \\                 
\hline 
\endfirsthead
\hline\hline
 & &   \\ 
\hline
\endhead
\hline
\endfoot
\hline
\multicolumn{3}{@{}l@{}}{\hbox to 0pt{\parbox{140mm}{\footnotesize
\footnotemark[$*$] Notations are described in the text.
\par\noindent
}\hss}}
\endlastfoot
\multicolumn{1}{l}{$\delta$}              & \multicolumn{2}{c}{$15$}\\
\multicolumn{1}{l}{$R$ ($\rm{cm}$)}       & \multicolumn{2}{c}{$7.0 \times 10^{16}$}\\
\multicolumn{1}{l}{$B$ ($\rm{Gauss}$)}    & \multicolumn{2}{c}{$0.71$}\\
\multicolumn{1}{l}{$\gamma_{\rm{min}}$}   & \multicolumn{2}{c}{$70$}\\
\multicolumn{1}{l}{$\gamma_{\rm{break}}$} & \multicolumn{2}{c}{$700$}\\
$\gamma_{\rm{max}}$   & $3300$               & $4500$ \\
$p$                   & $2.3$                & $2.0$ \\
$u_{\rm m}$ ($\rm{erg~cm^{-3}}$) & $2.0 \times 10^{-2}$ & $2.0 \times 10^{-2}$ \\
$u_{\rm e}$ ($\rm{erg~cm^{-3}}$) & $1.5 \times 10^{-3}$ & $2.1 \times 10^{-3}$ \\ 
\end{longtable}

\begin{figure}[ht]
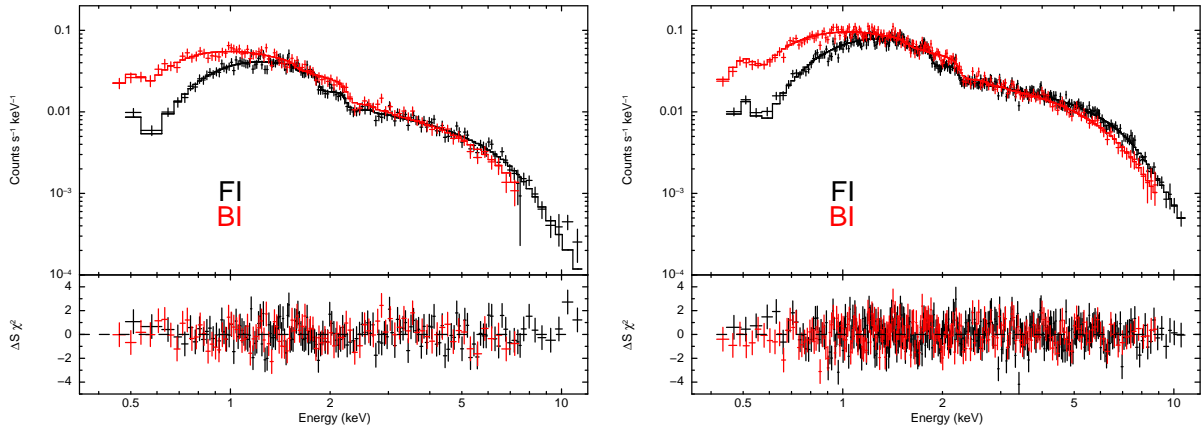

\begin{center}
\FigureFile(80mm,80mm){figure3a.ps}
\FigureFile(80mm,80mm){figure3b.ps}
\end{center}
\caption{
The XIS spectra of OJ~287 in MWL~I and MWL~II.
The FI and BI data are shown with black and red points, respectively.
The data are binned into pixels with at least $100$ events,
and error bars represent $1~\sigma$ statistical errors.
Histograms in both panels indicate the best-fit PL models.
}
\label{fig:XIS_spec}
\end{figure}

\begin{figure}[ht]
\begin{center}
\FigureFile(80mm,80mm){figure5a.ps}
\FigureFile(80mm,80mm){figure5b.ps}
\end{center}
\caption{
The HXD-PIN spectra of OJ~287 in MWL~II,
presented in the same manner as in figure~\ref{fig:PIN_qui}.
}
\label{fig:PIN_flare}
\end{figure}

\begin{thebibliography}{}
\bibitem[Abdo et al.(2009)]{fermi_detect}
        Abdo, A.~A., et al. 2009, arXiv:0902.1559 
\bibitem[Albert et al.(2007)]{BLLac}
	Albert, J., et al. 2007, \apj, 666, L17
\bibitem[Albert et al.(2008a)]{MAGICCrab}
	Albert, J., et al. 2008a, \apj, 674, 1037
\bibitem[Albert et al.(2008b)]{MAGICRF}
	Albert, J., et al. 2008b, Nucl. Instrum. and Meth., A588, 424
\bibitem[Aliu et al.(2009)]{MAGICtime} 
	Aliu, E., et al.\ 2009, Astroparticle Physics, 30, 293 
\bibitem[Bessell(1979)]{Bessell} 
	Bessell, M.~S.\ 1979, \pasp, 91, 589 
\bibitem[Boldt(1987)]{bolt} 
	Boldt, E.\ 1987, Observational Cosmology, 124, 611 
\bibitem[Capetti et al.(2002)]{Capetti} 
        Capetti, A., Celotti, A., Chiaberge, M., de Ruiter, H.~R., Fanti, R., Morganti, R., \& Parma, P.\ 2002, \aap, 383, 104 
\bibitem[Ciprini et al.(2008)]{Ciprini}
	Ciprini S.,et al. 2008, in proceedings of Workshop on Blazar Variability
	across the Electromagnetic Spectrum, PoS(BLAZARS2008)030 
\bibitem[Fiorucci \& Tosti(1996)]{F_and_T} 
	Fiorucci, M., \& Tosti, G.\ 1996, \aaps, 116, 403 
\bibitem[Fossati et al.(1998)]{Fos98}
        Fossati, G., Maraschi, L., Celotti, A., Comastri, A., \& Ghisellini, G. 1998, \mnras, 299, 433	
\bibitem[Fomin et al.(1994)]{wobble}
	Fomin, V. P., Stepanian, A., A., Lamb, R. C., Lewis, D. A., Punch, M., \& Weekes, T. C. 1994, Astropart. Phys., 2, 137 
\bibitem[Fukazawa et al.(2009)]{Fukazawa}
	Fukazawa, Y., et al. 2009, \pasj, 61, 17 
\bibitem[Ghisellini et al.(1998)]{Ghi98}
        Ghisellini, G., Celotti, A., Fossati, G., Maraschi, L., \& Comastri, A. 1998, \mnras, 301, 451
\bibitem[Giommi et al.(1999)]{S50716}
	Giommi, P., et al., 1999, \aap, 351, 59
\bibitem[Hillas(1985)]{Hillas}
	Hillas, A.~M. 1985, Proc. 29th Int. Cosmic Ray Conf. (La Jolla), 3, 445
\bibitem[Idesawa et al.(1997)]{Idesawa} 
	Idesawa, E., et al.\ 1997, \pasj, 49, 631
\bibitem[Ishida et al.(2007)]{XIS2HXD}
	Ishida et al. 2007, Suzaku Memo 2007-11
\bibitem[Ishisaki et al.(2007)]{Ishisaki} 
	Ishisaki, Y., et al.\ 2007, \pasj, 59, 113
\bibitem[Isobe et al.(2001)]{Isobe} 
	Isobe, N., Tashiro, M., Sugiho, M., \& Makishima, K.\ 2001, \pasj, 53, 79 
\bibitem[Inoue  \& Takahara(1996)]{Inoue} 
        Inoue, S., \& Takahara, F.\ 1996, \apj, 463, 555 
\bibitem[Kalberla et al.(2005)]{Kalberla} 
	Kalberla, P.~M.~W., Burton, W.~B., Hartmann, D., Arnal, E.~M., Bajaja,
	E., Morras, R., P{\"o}ppel, W.~G.~L.\ 2005, \aap, 440, 775
\bibitem[Kataoka(2000)]{Kataoka}
	Kataoka, J., Ph.D thesis, 2000, Univ. of Tokyo
\bibitem[Kidger(2000)]{Kidger} 
	Kidger, M.~R.\ 2000, \aj, 119, 2053 
\bibitem[Kokubun et al.(2007)]{Kokubun} 
	Kokubun, M., et al.\ 2007, \pasj, 59, 53 
\bibitem[Koyama et al.(2007)]{Koyama} 
	Koyama, K., et al.\ 2007, \pasj, 59, 23 
\bibitem[Kubo (1997)]{Kubo}
	Kubo, H., Ph.D thesis, 1997, Univ. of Tokyo
\bibitem[Li \& Ma(1983)]{LiMa}
	Li, T.-P., \& Ma, Y.-Q. 1983, \apj, 272, 317
\bibitem[Mitsuda et al.(2007)]{Mituda} 
	Mitsuda, K., et al.\ 2007, \pasj, 59, 1 
\bibitem[Okumura et al.(2000)]{Okumura}
	Okumura, S. K., et al. 2000, \pasj, 52, 393	
\bibitem[Padovani \& Giommi(1995)]{Pad95}
        Padovani, P. \& Giommi, P. 1995, \apj, 444, 567	
\bibitem[Pursimo et al.(2000)]{Pursimo}
        Pursimo, T., et al.\ 2000, \aaps, 146, 141 
\bibitem[Rolke et al.(2005)]{Rolke}
	Rolke, W. A., L¡­opez, A. M., \& Conrad, J. 2005, Nucl. Instrum. and Meth., A551, 493
\bibitem[Sambruna et al.(1994)]{EXOSAT}
        Sambruna, R.~M., Barr, P., Giommi, P., Maraschi, L., Tagliaferri, G., \& Treves, A.\ 1994, \apj, 434, 468 
\bibitem[Serlemitsos et al.(2007)]{Serlemitosos} 
	Serlemitsos, P.~J., et al.\ 2007, \pasj, 59, 9 
\bibitem[Shrader et al.(1998)]{OJ287_EGRET}
	Shrader, C.R., Hartman, R.C., Webb, J.R.,
	1996, \aap, 120, 599,
\bibitem[Skiff(2007)]{Skiff}
	Skiff, B. A., 2007, VizieR Online Data Catalog, 2277, 0
\bibitem[Skrutskie et al.(2006)]{Skrutskie}
	{Skrutskie}, M.~F., {Cutri}, R.~M., {Stiening}, R., {Weinberg}, M.~D.,
	{Schneider}, S., {Carpenter}, J.~M., {Beichman}, C., {Capps}, R., et al. 
	2006, \aj, 131, 1163-1183
\bibitem[Stickel et al.(1989)]{Stickel} 
	Stickel, M., Fried, J.~W., \& Kuehr, H.\ 1989, \aaps, 80, 103 
\bibitem[Sillanp\"{a}\"{a} et al.(1988)]{SMBH} 
	Sillanp\"{a}\"{a}, A., Haarala, S., Valtonen, M.~J., Sundelius, B., 
	\& Byrd, G.~G.\ 1988, \apj, 325, 628 
\bibitem[Sillanp\"{a}\"{a} et al.(1996b)]{double_peak}
	Sillanp\"{a}\"{a}, A., et al.\ 1996, \aap, 315, L13 
\bibitem[Sillanp\"{a}\"{a} et al.(1996a)]{OJ94} 
	Sillanp\"{a}\"{a}, A., et al.\ 1996, \aap, 305, L17 
\bibitem[Takahashi et al.(2000)]{Mrk421} 
	Takahashi, T., et al. 2000, \apj, 542, L105 
\bibitem[Takahashi et al.(2007)]{Takahashi} 
	Takahashi, T., et al.\ 2007, \pasj, 59, 35 	
\bibitem[Teshima et al.(2008)]{Tes08} 
	Teshima, M., et al.\ 2008, The Astronomer's Telegram, \#1500
\bibitem[Valtonen et al.(2006)]{Valtonen2} 
	Valtonen, M.~J., et al.\ 2006, \apjl, 643, L9 
\bibitem[Valtonen et al.(2008a)]{Valtonen_nature}
	Valtonen, M., et al., 2008, Nature, 452, 7189, 851
\bibitem[Valtonen et al.(2008b)]{OJ287_2005_2}
	Valtonen, M., Kidger, M.,  Lehto, H., \& Poyner, G.,
	2008, \aap, 477, 407
\bibitem[Valtaoja et al.(2000)]{disk-jet} 
	Valtaoja, E., Ter{\"a}sranta, H., Tornikoski, M., Sillanp{\"a}{\"a}, A.,
	Aller, M.~F., Aller, H.~D., \& Hughes, P.~A.\ 2000, \apj, 531, 744  
\bibitem[Madejski \& Schwartz(1988)]{Einstein} 
        Madejski, G.~M., \& Schwartz, D.~A.\ 1988, \apj, 330, 776 
\bibitem[Majumdar et al.(2005)]{MAGICMC}
	Majumdar, P., Moralejo, A., Bigongiari, C., Blanch, O., \& Sobczynska, D. 2005, in Proc. 29th Int. Cosmic Ray Conf. (Pune, India), 5, 203
\bibitem[Massaro et al.(2003)]{BeppoSAX} 
        Massaro, E., et al.\ 2003, \aap, 399, 33 
\bibitem[Watanabe et al.(2005)]{Watanabe}
	Watanabe, M., et al. 2005, \pasp, 117, 870
\end{thebibliography}
\end{document}